\documentclass[aps,pre,showpacs,manuscript]{revtex4}
\usepackage[centertags]{amsmath}
\usepackage{amssymb}
\usepackage{amsthm}
\usepackage{graphicx,subfigure}
\usepackage{epsfig}
\usepackage{multirow}
\usepackage{color}

\begin{document}

\title{Mechanism of barrier crossing dynamics in the presence of both time dependent and independent magnetic fields}

\author{Shrabani Mondal, Mousumi Biswas and Bidhan Chandra Bag
{\footnote{Author for correspondence,
e-mail:bidhanchandra.bag@visva-bharati.ac.in}}}

\affiliation{Department of Chemistry, Visva-Bharati, Santiniketan 731 235, India}

\begin{abstract}
In this paper we have presented the mechanism of the barrier crossing dynamics of a Brownian particle which is coupled to a thermal bath in the presence of both time independent and fluctuating magnetic fields. Here the following three aspects are important in addition to the role of the thermal bath on the barrier crossing dynamics. Magnetic field induced coupling may introduce a resonance like effect. Another role of the field is that enhancement of its strength reduces the frequency factor of the barrier crossing rate constant. Finally, the fluctuating magnetic field introduces an induced electric field which activates the Brownian particle to cross the energy barrier. As a result of interplay among these aspects versatile non-monotonic behavior may appear in the variation of the rate constant as a function of the strength of the time independent magnetic field. 
\end{abstract}


\maketitle
\newpage
{\bf Tuning of conductivity of solid electrolytes in the electrical devices is a challenging problem. Recently
it has been done in the experiment by changing the composition of the electrolytes. But it would be
very nice if there is a physical method to change the conductivity according to specific needs. The
Lorentz force may a good choice in this context. In the present study we have calculated the barrier crossing rate in the presence of both time dependent and independent
magnetic fields. It exhibits versatile non-monotonic behavior in the variation of the rate constant as a function of the strength of the time independent magnetic field. We have explored underlying mechanism of this observation. 
This work might be interesting in both theory and experiment.}
%\begin{multicols}{2}
%\newpage
\section{Introduction}
\noindent

Electrical devices play a key role for the modern civilization. The materials have potential applications in a diverse range of all-solid-state devices, such as rechargeable lithium batteries, flexible electrochromic displays and smart windows\cite{scros}. It may be necessary to tune the conductivity of the electrolytes which are present in the devices like these. For example, high ionic conductivity is needed for optimizing
the glassy electrolytes in various applications. The properties of the electrolytes are tuned by varying chemical
composition to a large extent and hence are adapted to specific
needs \cite{angel}. Thus the study of the ion conducting electrolytic materials is a very important area in physics and chemistry.
An alternative approach based on a physical method may be a challenging one in the context of tuning of properties of the electrolytes.  
One may explore the issue in the presence of the Lorentz force. Although time independent magnetic field (TIMF) cannot activate the particle to cross the barrier but it may effect the frequency of the dynamics. On the other hand time dependent magnetic field may introduce an induced electric field which can activate the particle. Another way is the direct application of an electric field which may be helpful in the case when very high rate of barrier crossing is necessary.
Thus using proper arrangement with the Lorentz force one may tune the conductivity of solid electrolytes according to specific need.  
In the very recent Refs. \cite{katsuki,pere,telang,vdo,bag3,bag4,aquino,bag5,bag6,physa}, it has been shown that the conductivity of an electrolytic material can be tuned by an applied magnetic field. The objective of the present study is related to this issue. Specifically,
we are interested to investigate about tuning of the barrier crossing rate with the change of strength of the time independent magnetic field in the presence of a fluctuating magnetic field. In this context we have considered a simple model that concerns the barrier crossing dynamics of a charged particle in the presence of both time independent and fluctuating magnetic fields along the $z$-direction. The barrier is introduced with double well potential along the $x$-direction. The particle may be bounded harmonically along the other directions. This model system has been studied using recently proposed Langevin equation of motion in the presence of the FMF\cite{physa}. Our investigation shows versatile non-monotonic behavior in the variation of the rate constant as a function of strength of the time independent magnetic field.
We have explored the origin of the non-monotonic behavior of the rate in detail. 

\noindent 
The outline of the paper is as follows: In Sec. II we have presented the model. The signature of the fluctuating magnetic fields on the barrier crossing dynamics has been explored in Sec.III. The paper is concluded in Sec. IV.

%\section{Calculation of the information
%entropy flux and production}

\section{The model}

In the present study we have considered the Brownian motion of a charged particle (having mass $m$) in the phase space in the presence of a magnetic field. 
%To make the study general we have considered that the applied magnetic field may be fluctuating one. 
The field is applied along $z$-direction i.e., $\textbf{B}_{app}=(0,0,B(t))$. The relevant Langevin equations of motion for this system can be written as follows\cite{bag6,physa};

\begin{equation}
\dot{x}=u_x\; \; \;, \label{eq1}
\end{equation}

\begin{equation}
\dot{y}=u_y\; \; \;, \label{eq2}
\end{equation}

\begin{equation}
\dot{u}_x =-4 a_0 x^3+2 b_0 x -\gamma_0 u_x +(\Omega_0+\Omega_f(t)) u_y+\frac{\dot{\Omega}_f(t) y}{2} + f_x(t)/m   \; \; \;, \label{eq4}
\end{equation}

\noindent
and

\begin{equation}
\dot{u}_y = -\omega^2  y-\gamma_0 u_y-(\Omega_0+\Omega_f(t)) u_x -\frac{\dot{\Omega}_f(t) x}{2} +f_y(t)/m \; \; \;.    \label{eq5}
\end{equation}
\noindent
In the above equations coordinate dependent conservative forces are derived from the following functional form for the potential energy,
$V(x,y)=m(a_0 x^4-b_0 x^2+\omega^2 y^2/2)$. The parameters,
$a_0$ and $b_0$ are constants which determine the fixed points corresponding to the potential energy function as well as the energy barrier height. The remaining parameter, $\omega$ is the characteristic frequency of the harmonic oscillation along $y$ direction. The coordinates of the minima of the potential function are $(-\sqrt{\frac{b_0}{2a_0}},0)$ and $(\sqrt{\frac{b_0}{2a_0}},0)$. The maximum of the potential energy function is located at the coordinate, (0,0). Thus the energy barrier is $\frac{mb_0^2}{4a_0}$. The parameter $b_0$ has another significance.
It is related to the frequencies of the motion along $x$-direction around the bottom and the top of the potential energy function. The angular frequency, $\omega_a$ at the bottom can be read in terms of $b_0$ as $\omega_a=2\sqrt{b_0}$. Similarly the angular frequency, $\omega_b$ at the barrier top one may read as $\omega_b=\sqrt{2b_0}$. Shortly we will demonstrate that $\omega$, $\omega_0$ and $\omega_b$ are related to the barrier crossing frequency. In other words,
these parameters characterize the magnitude of the position dependent force due to the electrostatic interaction of the concerned charged particle with the particles which are surrounding it. Thus their magnitudes may depend on the characteristics of all the charged particles which are associated with the system and its environment. Hence the energy barrier as well the frequency of the hopping dynamics are governed by the composition of the electrolyte.  

   Another deterministic force in the Langevin equations of motion is due to the environment induced dissipative force which depends on the velocity of the particle and the damping strength, $\gamma_0$. It considers the flow of the mechanical energy from the system  to the surroundings as a signature of the unavoidable coupling \cite{physa} between these two. The coupling strength\cite{physa} certainly depends on the nature of the interaction between the system and the constituents in the surroundings. Thus $\gamma_0$ also may depend on the composition of the electrolyte. Now we would mention here that the system experiences a random force with components $f_x$, $f_y$ and $f_z$ from the constituents in the surroundings as a result of their thermal motion. We assume that it is Gaussian in nature. The components ($f_x$ and $f_y$) of the random force in the Langevin equations of motion are independent and related with the damping strength by the standard fluctuation-dissipation relation,
$\langle f_x(t) f_x(t')\rangle=  \langle f_y(t)f_y(t')\rangle=\langle f_z(t)f_z(t')\rangle=2 \gamma_0 mk_BT\delta(t-t')$. It is to be noted here that $f_z$ is not a relevant one in the present case as we are interested to study the motion in the $x-y$ plane which experiences the  forces from the applied magnetic field and the induced electric field from it, respectively. However, $\gamma_0$ has an important significance in the dynamics. It reduces the frequency which may be associated with dynamics. Therefore the barrier crossing frequency may depend on the damping strength\cite{bag3}. The activation energy also may depend on $\gamma_0$\cite{spre} if the stationary state is not a Boltzmann one since the probability at 
top of the energy barrier depends on the energy dissipation rate at the steady state. Thus $\gamma_0$ has dual role in the present study to determine the barrier crossing rate at long time as the time dependent magnetic field produces a steady state instead of an equilibrium one.
We now mention the role of the temperature in the dynamics. The fluctuation dissipation relation implies that the strength of the random force depends on the temperature. Thus it measures the diffusion as well as the probability of the particle in space. The significance of temperature in the barrier crossing process is observed through the probability of the particle at the barrier top. But in sum cases where the multiplicative noise induced drift may depend on the temperature\cite{spre} then the frequency factor of the barrier crossing rate may depend on the temperature also. Here the activation energy may depend on the temperature\cite{spre}. At this circumstance the barrier crossing rate at stationary state may deviate from the Arrhenius law\cite{spre} due to these reasons. In the presence of the fluctuating magnetic field  may not be valid if the strength of the fluctuating magnetic field is relatively higher\cite{bag6}.

   We now consider the quantities $\Omega_0$ and $\Omega_f(t)$ in the magnetic force term. 
$z-$ component of the applied magnetic field can be split as,
$B(t)=B_0+\eta(t)$. 
Here $B_0$ is a constant part of the $B(t)$ and $\eta(t)$ is the fluctuating one. The splitting may be due to the fluctuation of current in the electrical magnet. Then the cyclotron
frequency ($\Omega$) becomes time dependent and it can be read as,
$\Omega= \Omega_0+\Omega_f (t)$ with $\Omega_0=\frac{qB_0}{m}$ and
$\Omega_f(t)=\frac{q\eta(t)}{m}$. It may modify the frequency which is derived from the curvature of the potential energy function. Shortly we will demonstrate this aspect for a special case. There one may find that the effective frequency of the dynamical system decreases with increase in strength of time independent magnetic field. It implies that the frequency factor of the barrier crossing rate constant reduces as the field becomes stronger. We now consider the another point.
The time dependent magnetic field induced electric field contributes a force. It appears in the Langevin equations of motion after the magnetic force term. Now one may ask why fluctuations in $B(t)$ are restricted to the $z$ direction? The form of $\textbf{B}_{app}$ has been chosen keeping in mind the objective of the present study. In the Langevin equations of motion we have considered that the Brownian particle may experience a finite energy barrier along the $x$ direction. Then to investigate the role of the magnetic field we would apply the field at least along one of the perpendicular directions to this direction. Keeping it in mind we have considered the conventional simple choice about the applied magnetic field which is directed along the $z$ direction. Through this one may explore the role of the velocity dependent magnetic force and the induced electric field (which is due to time dependent magnetic field) on the barrier crossing dynamics in the four dimensional phase space
as suggested by the above set of the Langevin equations of motion. Further addition of the field along $x$ or $y$ direction or in  both directions does not include new kind of force but increases two more dimensions in the phase space description. These are the probable reasons regarding the choice about the applied magnetic field. The electric field which is derived from it helps to activate the Brownian particle to cross the energy barrier. Shortly we will consider this aspect in detail.  

   Regarding the equations of motion another point would be noted here. Eqs.(\ref{eq4}-\ref{eq5}) look like incomplete description of the motion since these do not include the effect from the induced magnetic field due to the time dependent electric field as suggested by the Maxwell's equation. One may easily check here that the induced electric field which appears in the equation of the motion is related to the applied magnetic field by
   
\begin{equation}
\nabla \times \textbf{E}_{ind}=-\frac{\partial }{\partial t}\textbf{B}_{app}=-\dot{\eta}(t) \; \; \;, \label{indel1}
\end{equation}

\noindent
where $\textbf{E}_{ind}(=\left(\frac{\dot{\eta}(t) y}{2},-\frac{\dot{\eta}(t) x}{2},0\right))$ is the induced electric field. The Maxwell's equation with this electric field for the present system can be read as

\begin{equation}
\nabla \times \textbf{B}_{ind}=\mu\epsilon\frac{\partial }{\partial t}\textbf{E}_{ind}  \; \; \;. \label{indmf}
\end{equation}

\noindent
The applied magnetic field does not appear in the above equation since its curl is zero. $\epsilon$ and $\mu$ correspond to the permitivity and
the permeability of the electrolytic medium. Taking curl in the both sides of the above equation and then using Eq.(\ref{indel1}) and  
$\nabla \bf{\cdot} \textbf{B}_{ind}=0$ into this we have

\begin{equation}
\frac{\partial^2 }{\partial^2 z}B_{indz}=\mu\epsilon\ddot{\eta}(t)  \; \; \;, \label{indmf1}
\end{equation} 

\noindent
where $B_{indz}$ is the $z$-component of the induced magnetic field. The solution of the above equation can be read as

\begin{equation}
B_{indz}=\mu\epsilon\ddot{\eta}(t)z^2 =\mu\epsilon z^2\frac{\partial^2 }{\partial t^2}\textbf{B}_{app}\; \; \;. \label{indmf2}
\end{equation} 

\noindent
Here we have used $\frac{\partial }{\partial z}B_{indz}=0$ at $t=0$ and $z=0$. It is to be noted here that 
the magnitude of product of $\mu$ and $\epsilon$ may be of the order of $10^{-17} (m/S)^{-2}$. Then above equation certainly implies that the induced magnetic field is negligible compared to the applied magnetic field. This might be the reason to exclude the effect from the induced magnetic field in the equation of motion in the earlier studies\cite{bag5,bag6,physa,indmag}.

 To broaden the scope of the present study we assume that $\eta(t)$ is a colored noise. The Ornstein-Uhlenbeck  noise usually has been considered in the literature\cite{bagmal,hang1} to capture the essential features of the non-Markovian dynamics. In the present study, we have also considered that $\eta(t)$ is the Ornstein-Uhlenbeck  noise. Then its time evolution equation is given by

\begin{equation}\label{eq12}
\dot{\eta} =  -\frac{\eta}{\tau}+\frac{\sqrt{D}}{\tau} \zeta(t) \; \;.
\end{equation}

\noindent
$D$, in the above equation measures the strength of the fluctuating magnetic field and $\tau$ is the correlation time. $\zeta(t)$  in Eq.(\ref{eq12}) corresponds to a white Gaussian noise having variance two. The two time correlation function of the fluctuating field
can be read as, $\langle \eta(t) \eta(t') \rangle = \frac{D}{\tau} e^\frac{-|t-t'|}{\tau}$. Based on the above equation the time evolution of the fluctuating Larmor frequency can be read as

\begin{equation}\label{eq12f}
\dot{\Omega}_f =  -\frac{\Omega_f}{\tau}+\frac{q\sqrt{D}}{m\tau} \zeta(t) \; \;.
\end{equation}

Now using the above  equation into Eqs.(\ref{eq4}-\ref{eq5}) one can write that

\begin{equation}
\dot{u}_x =-4 a_0 x^3+2 b_0 x -\gamma_0 u_x +(\Omega_0+\Omega_f(t)) u_y-\frac{\Omega_f(t) y}{2\tau}+\frac{q\sqrt{D}y}{2m\tau} \zeta(t)
 + f_x(t)/m   \; \; \;, \label{eq14}
\end{equation}

\noindent
and

\begin{equation}
\dot{u}_y = -\omega^2  y-\gamma_0 u_y-(\Omega_0+\Omega_f(t)) u_x +\frac{\Omega_f(t) x}{2\tau}-\frac{q\sqrt{D}y}{2m\tau} \zeta(t) +f_y(t)/m \; \; \;.    \label{eq15}
\end{equation}

We now mention the relevance of the Langevin Eqs.(\ref{eq1}-\ref{eq5}). 
Thermally activated ion hopping in liquid \cite{rob,con} and solid \cite{strom,angel,christie,muru} is well known. These are examples of the Kramers' problem\cite{kram,hangimoj,marche,revm,melni,revm1}. Still active research is going on this issue\cite{bag3,bag4,aquino,bag5,bag6,physa,jsm,alendu_jcp,poll3,poll5,tiwar,jsm1,chaos,chaos1}. 
Ion moves to the neighboring site along the direction of transport through hopping. Two consecutive sites may be represented by the double well potential assuming that ion is transported along $x$-direction. The barrier height of the well mimics the potential energy barrier which is experienced by an ion during its hopping to the next site. Here one may effectively assume that motion of ion is harmonically bound along $y$ and $z$ directions. Under this circumstance if there is a  magnetic field along $z$-direction from an electrical magnet (the field due to it may be fluctuating one)  then the ion may mimic the dynamics as described
by Eqs.(\ref{eq1}-\ref{eq5}). Similarly the above equations of motion may find application in the case of movement of ion in solid.

Another point is to be noted here that
the presence of the fluctuating magnetic fields as well as the induced electric field leads to an unusual type open system (for further details we refer our recent work \cite{bag6}) by virtue of the velocity and  the coordinates dependent multiplicative noises, respectively.  Thus it is difficult to solve the coupled Langevin equations (\ref{eq1}-\ref{eq2}, \ref{eq12}, \ref{eq14}-\ref{eq15})  as well as to write the Fokker-Planck equation in the phase space. 
Here the fluctuating part, $\eta(t)$ of the total magnetic field creates major problem to investigate the 
present study analytically. For $\eta(t)=0$, one can solve the above coupled equations of motion following Ref.\cite{aquino1}. Another point is to be noted here that it is difficult to invoke the unified colored noise approximation \cite{hang,jia1,hang1,cao} to reduce the dimension of the problem even at over damped limit. Ignoring the effect of inertia and thermal noise on the dynamics  one may solve the equations of motion as shown in the Appendix. But calculation of relevant moments based on this solution is not simple even applying further approximation since cosine and exponential functions appear in the solution as the very complex functions of noises $\eta(t)$ and $\zeta(t)$, respectively.   

Even the study of additive colored noise driven barrier crossing dynamics in the presence of a time independent magnetic field is not simple. Very recently\cite{aquino} the barrier crossing rate has been calculated at quasi deterministic limit for the additive colored fluctuating electric field driven process in the presence of the TIMF. However, in the recent past, study on the Brownian motion of the charged particle in the presence of a magnetic field has been considered in different contexts. But most of them  have used the Markovian description of the motion. It does not mean that the non-Markovian dynamics of charged particle is not important in nature. There are several examples where the Brownian motion of the charged particle is found to occur in the condensed phase\cite{angel,rog,levine,rob,hsieh,vice}. 
It is to be noted here that although the study of the Brownian motion of charged particle in the presence of the magnetic field  is getting strong attention in different contexts \cite{katsuki,pere,telang,vdo,bag3,bag4,aquino,bag5,bag6,physa,pnik,mar,fabio,jayn,gelf,nmar,nmar1,nmar3,nmar4} but 
till now few studies have been reported considering the non-Markovian dynamics \cite{bag4,bag6,nmar,nmar1,aquino,nmar3,nmar4}. Most of these studies with non-Markovian dynamics have been done based on the Langevin equation of motion. In Refs.\cite{bag4,nmar,nmar1}, the theoretical formulation with the Fokker-Planck equation in the extended phase space (the noise as one of the phase space variables) for the non-Markovian thermal bath is given with the
Ornstein-Uhlenbeck thermal noise. 
Very recently the Fokker-Planck equation has been proposed in the natural phase space in a simple way for arbitrary frictional memory kernel\cite{jcp} and additive thermal noise. The Fokker-Planck equation in the phase space also has been derived in a recent Ref.\cite{nmar4}
using a characteristic function for a Brownian charged particle embedded in a memory thermal bath and under the action of force fields: a constant magnetic field and arbitrary time-dependent force fields. 
Based on these methods it is very difficult to write the required Fokker-Planck equations in the energy space or the phase space for the present system where both coordinate and velocity dependent colored multiplicative noises appear. 
Thus we are restricted to study the present problem numerically with fluctuating magnetic field as given by Eqs.(\ref{eq14}-\ref{eq15}) to identify its special role (if any). In the following sections we will present our observations.

\section{Non-Monotonic functional behavior of the barrier crossing rate in terms of strength of the time independent magnetic field}

To study the present problem we have solved Eqs.(\ref{eq1}-\ref{eq2}, \ref{eq12f},\ref{eq14}-\ref{eq15}) numerically using the Heun's
method \cite{rt}. Based on this method we calculate the first passage time $t_f$, that is the time required for a trajectory which starts from the coordinate ($x=-\sqrt{b/2a}, y=0$) corresponding to the left minimum (most probable state) of the two-dimensional potential to reach the top of the energy barrier($x=0, y=0$) for the first time. For further details in this context we refer to Ref.\cite{physa} where the numerical calculation procedure has been discussed in depth. The reliability of the method is well justified in Ref.\cite{bag6}. However, the barrier crossing is assisted by the noise and therefore $t_f$ is a statistical quantity. We have determined its mean value, $<t_f>$, over many realizations of the trajectories such that the mean becomes independent of the number of realizations. In general, we have considered 20000 to 25000 trajectories for the statistical averaging. Inverse of the mean first passage time gives the barrier crossing rate constant, $k=\frac{1}{<t_f>}$. 

\begin{figure}[!htb]
\includegraphics[width=16cm,angle=0,clip]{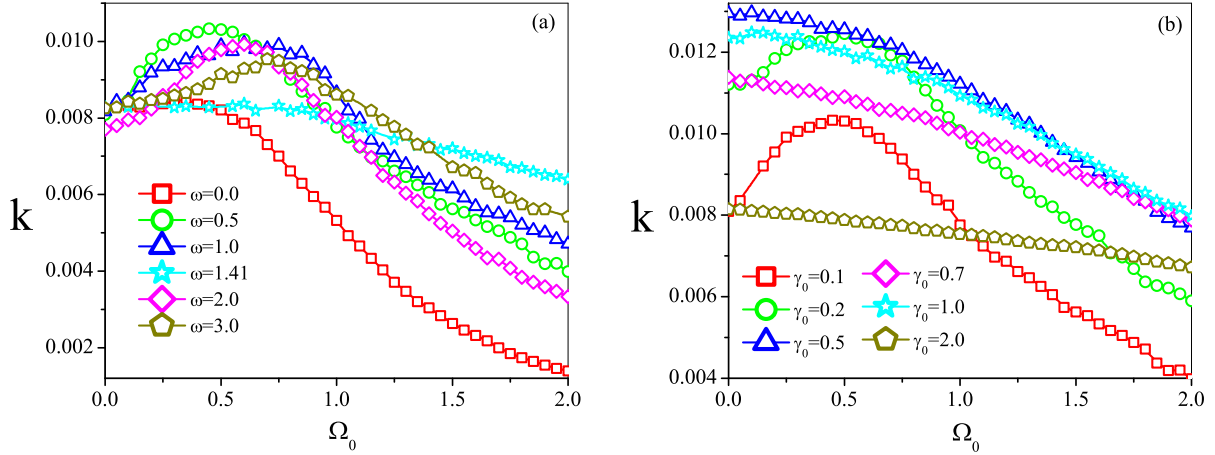}
\caption{(Color online) Demonstration of the resonance effect induced turnover phenomenon. Plot of the rate constant, $k$ {\it vs.} strength of the time independent magnetic field for the common parameter set, $a_0=0.25$, $b_0=0.5$, $ \frac{k_BT}{m}=0.1$ and $\frac{q\sqrt{D}}{m}=0.0 $.  (a) $\gamma_0=0.1 $ (b) $\omega=1.0 $. (Units are arbitrary.)}
\label{fig1}
\end{figure}

To begin our investigation we have calculated the barrier crossing rate constant as a function of strength of the time independent magnetic field in the absence of fluctuating magnetic field and presented in Fig.1. 
It exhibits that at low damping limit, if the frequencies around the fixed points of the double well oscillator is close to frequency of the harmonic motion along the $y$ direction, then there is an optimum behavior in the variation of the rate constant as a function of strength of TIMF. Here the rate passes through a maximum. But the maximum may disappear at relatively large damping strength as shown in Fig.1(b). Then the turnover phenomenon seems to be a resonance induced one. The magnetic field induced coupling may introduce the resonance between the two directions at low damping regime if the frequency of the harmonic oscillator along the y-direction is close to the frequency of either bottom or top of the double well. The effective frequency for  each component of motion can be estimated at a limiting situation i.e. , $\gamma_0\rightarrow 0$. In this context we also consider harmonic potential along $x$ direction. Then the equations of motion (\ref{eq14}-\ref{eq15}) in the absence of both the thermal noise and the fluctuating magnetic field become 
 
\begin{equation}
m\dot{u}_x = -\omega^2 m x -\gamma_0 mu_x +qB_0 u_y \label{eq33d}
\end{equation}

\noindent
and

\begin{equation}
m\dot{u}_y = -\omega^2 m y-\gamma_0 m u_y-qB_0 u_x   \; \; \;. \label{eq34d}
\end{equation}

\noindent
It is to be noted here that in Eq.(\ref{eq33d}) we have considered the dynamics around the bottom of the double well with angular frequency, $\omega_a(=2\sqrt{b_0}=\omega)$. The above coupled equations of motion can be solved using the transformation, $\xi=x+iy$\cite{aquino1,landau}. Then we have

\begin{equation}
\ddot{\xi} = -\omega^2 \xi -2\beta\dot{\xi} \label{eqzeta} \; \; \;,
\end{equation}

\noindent
where $\beta=\frac{\gamma_0 +i\Omega_0}{2}$. This leads to have the solution of the above equation as

\begin{equation}
\xi(t) = \xi(0)e^{-\beta t} cos(\sqrt{\omega^2-\beta^2}t) \label{eqzeta1} \; \; \;.
\end{equation}

\noindent
Decomposing the above solution at the limit $\gamma_0\rightarrow 0$, time dependence of $x(t)$ and $y(t)$ can be read as

\begin{eqnarray}
x(t) & = & \frac{x(0)}{2}[cos(\sqrt{\omega^2+\Omega_0^2/4}+\Omega_0/2)t+cos(\sqrt{\omega^2+\Omega_0^2/4}-\Omega_0/2)t]\nonumber\\
&+&\frac{y(0)}{2}[sin(\sqrt{\omega^2+\Omega_0^2/4}+\Omega_0/2)t-sin(\sqrt{\omega^2+\Omega_0^2/4}-\Omega_0/2)t]
\label{eqzeta4} \; \; \;
\end{eqnarray}

\noindent
and

\begin{eqnarray}
y(t) & = & \frac{y(0)}{2}[cos(\sqrt{\omega^2+\Omega_0^2/4}+\Omega_0/2)t+cos(\sqrt{\omega^2+\Omega_0^2/4}-\Omega_0/2)t]\nonumber\\
&-&\frac{x(0)}{2}[sin(\sqrt{\omega^2+\Omega_0^2/4}+\Omega_0/2)t-sin(\sqrt{\omega^2+\Omega_0^2/4}-\Omega_0)/2)t]
\label{eqzeta5} \; \; \;.
\end{eqnarray}

\noindent
One can now easily check that the above relations reduce to the expected results at the limit $\Omega_0=0.0$. However, $x(t)$ as well as $y(t)$ are superposition of periodic terms with periods $T_1=\frac{2\pi}{\sqrt{\omega^2+\Omega_0^2/4}+\Omega_0/2}$ and $T_2=\frac{2\pi}{\sqrt{\omega^2+\Omega_0^2/4}-\Omega_0/2}$ and their ratio can be read as $\frac{T_2}{T_1}=\frac{\omega^2+\Omega_0^2/2+\Omega_0\sqrt{\omega^2+\Omega_0^2/4}}{\omega^2}$. Under this circumstance, a resonance phenomenon may appear between the two directions. Another important point is to be noted here. Increase in $T_2$
with $\Omega_0$ implies that the frequency of the system decreases as the strength of the time independent magnetic field grows. It suggests that the frequency factor of the barrier crossing rate would decrease with increase in $\Omega_0$. Thus the TIMF plays dual role as well as introduces two opposite effects on the barrier crossing rate. As a result of that a turnover may appear. It is to be noted here that Fig.1 is quite similar to Fig.5 in Ref.\cite{physa}. To account for the crossover phenomenon (which may appear in the 
plot the rate constant {\it vs.} damping strength ($\gamma_0$) as shown in Fig.1 in Ref.\cite{physa} for different $\Omega_0$) the Fig.5 was considered with very qualitative arguments. The turnover as described in Fig.1 was reported first in Ref.\cite{bag3}. But the above discussion was missing in earlier Refs.\cite{bag3,bag5,physa}. The barrier crossing rate has been calculated quantitatively in Refs.\cite{bag3,bag4} at intermediate value to strong damping regime to investigate the role of the time independent magnetic field. Here it has been shown that the frequency factor of the barrier crossing rate decreases with increase in the strength of the TIMF. We now summarize this observation to make the presentation self-sufficient.
In the presence of the TIMF the energy of an isolated system is conserved. Therefore the phase space distribution function at equilibrium is a Boltzmann one for a Brownian particle which is coupled to the thermal bath in the presence of the time independent magnetic field\cite{bag3}. It implies that the frequency factor of the barrier crossing rate may influence by the field. This can be read for the Langevin equations of motion(\ref{eq14}-\ref{eq15}) with $\eta(t)=0$ as \cite{bag3} 

\begin{equation}
k  =\frac{\omega_a}{2\pi}\sqrt{\frac{\lambda \gamma_0^2}{(1+a_5^2) r_1 r_2 r_3}} \exp \left(-\frac{E_b}{k_BT}\right) \; \;,
\label{eqk0}
\end{equation}

\noindent
where
$E_B=\frac{mb_0^2}{4a_0}$, $r_1 =\frac{\gamma_0(1+a_5^2)+\lambda}{(1+a_5^2)}$, $r_2 =\frac{(1+a_5^2)(\gamma_0^2+\lambda\gamma_0}{\gamma_0(1+a_5^2)+\lambda}$,
$r_3=\frac{\lambda\gamma_0(1+a_5^2)+a_5^2\omega^2}{\lambda(1+a_5^2)}-\frac{a_5^2\omega^2}{(1+a_5^2)(\gamma_0(1+a_5^2)+\lambda} 
-\frac{a_5^4\omega^2\gamma_0^2}{(\gamma_0(1+a_5^2)+\lambda)^2 r_2}$,
$a_5  =- \frac{\lambda \Omega_0}{\lambda^2+\gamma_0 \lambda+\omega^2}$ and 
$\lambda$ is the solution of the algebraic equation,
$\lambda^4+2\gamma_0\lambda^3 +(\Omega_0^2+\omega^2-w_b^2 
+\gamma_0^2)\lambda^2+\gamma_0(\omega^2-\omega_b^2) \lambda-\omega^2\omega_b^2 =0$. Here $w_b(=\sqrt{2b_0})$ is the angular frequency at the energy barrier of the double well potential.
The above expression for the rate constant suggests that frequency factor of the barrier crossing rate decreases with increase in
the strength of the constant magnetic field.  We have addressed origin of this fact above through the solution of the equations (\ref{eq33d}-\ref{eq34d})
of motion. To avoid any confusion it is to be noted that the turnover as demonstrated in Fig.1(a) cannot be described by Eq.(\ref{eqk0})
since it is valid in the regime, intermediate to high damping strength.

\begin{figure}[!htb]
\includegraphics[width=16cm,angle=0,clip]{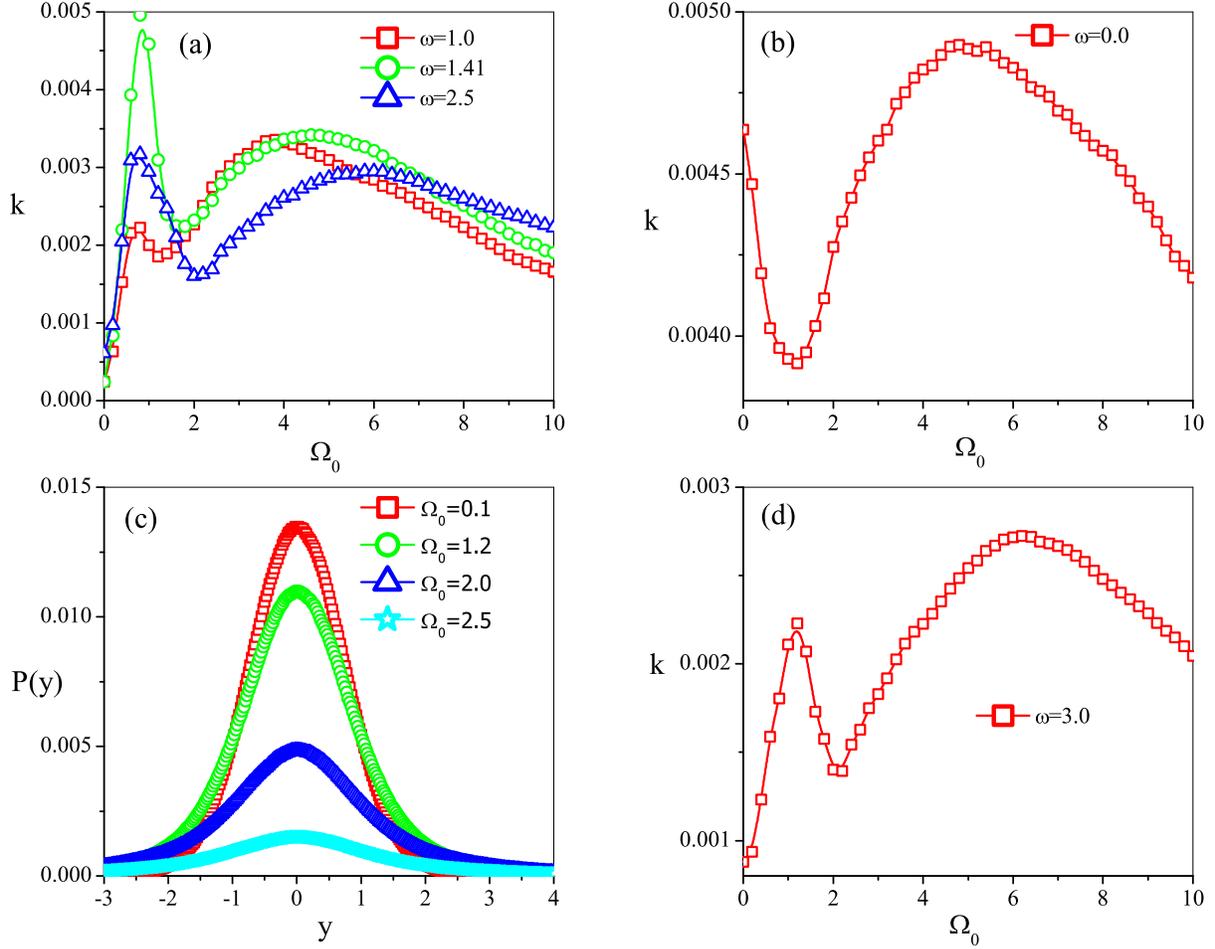}
\caption{(Color online) Fluctuating magnetic field induced tri-turnover phenomenon. The common parameter set for this figure is $a_0=0.25$, $b_0=0.5 $,  $\gamma_0=0.1$, $\frac{k_BT}{m}=0.025$, $\frac{q\sqrt{D}}{m}=0.14$ and $\tau=0.1$. (a) 
Plot of the rate constant, $k$ {\it vs.} strength of the time independent magnetic field.
(b) Plot of the rate constant, $k$ {\it vs.} strength of the time independent magnetic field for $\omega=0$. (c) Plot of reduced distribution function, $P(y)$ {\it vs.} $y$ for the same parameter set as in (b). (d) 
Plot of the rate constant, $k$ {\it vs.} strength of the time independent magnetic field for $\omega=3.0$. (Units are arbitrary.)}
\label{fig2}
\end{figure}

We now explore the fate of the non-monotonic behavior of the rate constant in the presence of a fluctuating magnetic field. In this context we have demonstrated the variation of the rate constant as a function of the strength of the TIMF in Fig.2. It shows that the FMF may induce
a tri-turnover phenomenon. Thus non-monotonic behavior becomes more complex in the presence of the fluctuating magnetic field.
In this context panel (b) in Fig.2 is quite interesting. It exhibits that 
in the absence of the harmonic force along y direction the tri-turnover phenomenon may become bi-turnover one.
Then this panel along with the Fig.1(a) clearly suggests that the first turnover in Fig.2(a) is a signature of the interplay between the resonance effect and the decrease of the barrier crossing frequency factor with increase in the strength of the constant magnetic field.
One may now ask how the second turnover appears? There is an another obvious question as well as doubt about the panel (b) of Fig.2. In the absence of the harmonic force along the y direction, the motion in this direction as well as another direction may be unbounded by virtue of coordinate dependent coupling of equations of motion. Under this circumstance how would the Fig.2(b) be possible. To have the answer to these questions we have calculated the reduced distribution function of position for the the y-direction motion and demonstrated in panel (c) of  Fig.2. It is apparent in this panel that the motion along y-direction is effectively harmonically bounded by virtue of coupled equations of motion due to the magnetic force and the induced electric field, respectively. To describe it mathematically we consider harmonic potential(with angular frequency, $\omega$) along both the directions. Then the effective equation of motion along the individual direction can be read as \cite{bag6,physa}          

\begin{eqnarray}\label{bn1}
\dot{u}_x & = & -c'(t) x -\gamma'(t) u_x+(\Omega_0+\Omega_f(t))[\frac{d}{dt}(e^{-Kt}\alpha(t))-\Omega_0\frac{d}{dt}(L^{-1}u_x) \nonumber\\
& - &\Omega_f(t)\frac{d}{dt}(L^{-1}u_x) -(L^{-1}u_x)\frac{d}{dt}\Omega_f(t)
-\frac{\dot{\Omega}_f(t)(t)}{2}\frac{d}{dt}(L^{-1}x)-\frac{(L^{-1}x)}{2}\frac{d}{dt}\dot{\Omega}_f(t)(t)\nonumber\\
&+&\frac{d}{dt}(\frac{L^{-1}f_y(t)}{m})] +\frac{\dot{\Omega}_f(t)}{2}[e^{-Kt}\alpha_1(t)-\Omega_0 L^{-1}u_x-\Omega_f(t)L^{-1}u_x  \nonumber\\
& - &\frac{\dot{\Omega}_f(t)}{2}L^{-1}x+ \frac{L^{-1}f_y(t)}{m}]+\frac{f_x(t)}{m} 
\end{eqnarray}

\noindent
and

\begin{eqnarray}\label{bn2}
\dot{u}_y & = & -c'(t) y -\gamma'(t) u_y-(\Omega_0+\Omega_f(t))[\frac{d}{dt}(e^{-Kt}\alpha(t))+\Omega_0\frac{d}{dt}(L^{-1}u_y) \nonumber\\
& + &\Omega_f(t)\frac{d}{dt}(L^{-1}u_y) +(L^{-1}u_y)\frac{d}{dt}\Omega_f(t)
+\frac{\dot{\Omega}_f(t)(t)}{2m}\frac{d}{dt}(L^{-1}y)+\frac{(L^{-1}y)}{2}\frac{d}{dt}\dot{\Omega}_f(t)(t)\nonumber\\
&+&\frac{d}{dt}(\frac{L^{-1}f_x(t)}{m})]-\frac{\dot{\Omega}_f(t)}{2}[e^{-Kt}\alpha_1(t)+\Omega_0 L^{-1}u_y+\Omega_f(t)L^{-1}u_y  \nonumber\\
& + &\frac{\dot{\Omega}_f(t)}{2}L^{-1}y+\frac{L^{-1}f_x(t)}{m}]+\frac{f_y(t)}{m} \; \; \;,
\end{eqnarray}

\noindent
where  $e^{-Kt}\alpha(t)$ with $K=\frac{\gamma_0}{2}$ is the solution of Eq.(\ref{eq33d}) at the limit, $B_0=0.0$ and $L=\frac{d^2}{dt^2}+\gamma_0 \frac{d}{dt}+\omega^2$. The other quantities are defined as

\begin{equation}\label{bn3}
 c'(t)=\omega^2+\frac{\dot{\Omega}_f(t)}{4}L^{-1}\dot{\Omega}_f(t)+\frac{(\Omega_0+\Omega_f(t))}{2}\frac{d}{dt}(L^{-1}\dot{\Omega}_f(t))\; \; \;,
\end{equation}

\noindent
and

\begin{eqnarray}\label{bn4}
\gamma'(t)=\gamma_0 +\frac{\dot{\Omega}_f(t)}{2}L^{-1}\Omega_f(t)+(\Omega_0+\Omega_f(t))[\frac{d}{dt}(L^{-1}\Omega_f(t))+(L^{-1}\Omega_f(t))\frac{d}{dt} \
+\frac{L^{-1}\dot{\Omega}_f(t)(t)}{2}]\; \; \;.
\end{eqnarray}

\noindent
Thus the above equations  imply that the motion may be bounded harmonically even in the absence of the harmonic potential and the time independent magnetic field. We now come back to Fig.2(c). This figure shows that the width of the distribution function increases as the strength of the TIMF grows. It is difficult to explain this observation based on the above equations of motion. However,
in the initial part of this section we have shown that the effective frequency of the bounded motion decreases with increase in $\Omega_0$. Thus the damping effect may weaken due to the enhancement of the strength of the time independent magnetic field. Under this circumstance the probability of attaining of the large amplitude as well as width of the distribution function increases as shown in panel (c) of Fig.2. Then it is easy to explain both the second and the third turnovers in the panel (a) of Fig.2. According to Fig.2(c) the induced electric field becomes more effective to activate the particle to go to the energy barrier with the enhancement of the strength of the TIMF. Thus around the second turnover the electric field induced activation takes the leading role in the barrier crossing dynamics and it may continue until the third turnover appears. After the third turnover the decrease in the frequency factor of the rate constant  with increase in the strength of the TIMF again becomes important.

We now consider the last panel of Fig.2. It is also quite interesting.  
Here the first turnover  may not be expected according to Fig.1(a). Based on Eq.(\ref{bn3}) one may resolve the issue. The re-normalized (effective) frequency as suggested by Eq.(\ref{bn3}) is responsible to determine whether the resonance will occur or not. Thus anticipation based on Fig.1(a) may not be correct in the presence of a fluctuating magnetic field. In that case the tri-turnover phenomenon is  purely the fluctuating magnetic field induced one. Some other related points are to be noted here.
Fig.2 shows that the resonance effect is better for the case where the frequency of harmonic oscillator matches with the frequency at the bottom of the double well oscillator. Fig.2 also shows that  the second turnover starts at the higher strength of the time independent magnetic field with increase in the frequency of the harmonic oscillator. One may explain this considering a figure which is similar to Fig.2(c). For a given strength of the TIMF, the width of the distribution function is enhanced as the frequency of the harmonic oscillator decreases. Thus the induced electric field may be effective significantly in the barrier crossing dynamics even at lower strength of the time independent magnetic field for smaller frequency of the harmonic oscillator as implied in Fig.2.

\begin{figure}[!htb]
\includegraphics[width=16cm,angle=0,clip]{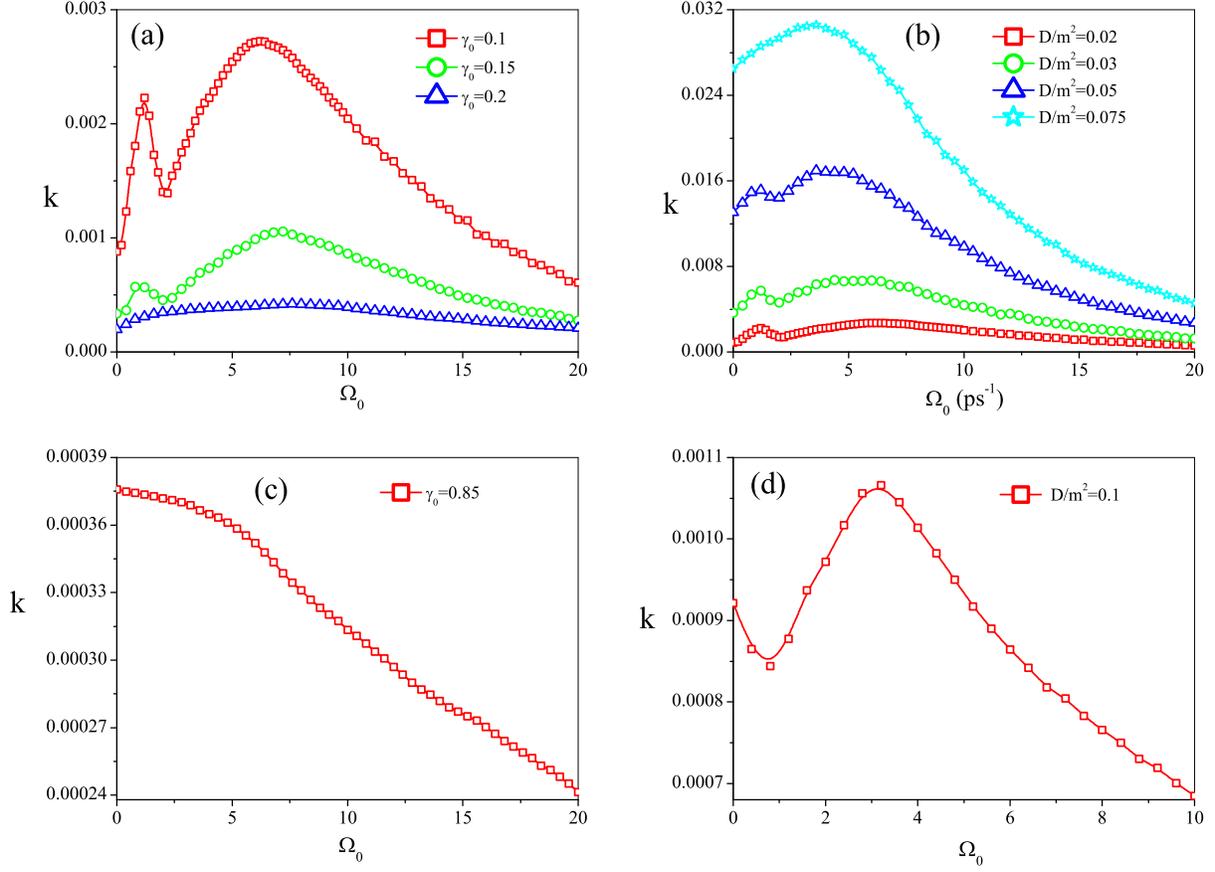}
\caption{(Color online) Conversion of tri turnover phenomenon into mono or bi-turnover phenomenon on changing strengths of the damping and the fluctuating magnetic field, respectively. 
Plot of the rate constant, $k$ {\it vs.} strength of the time independent magnetic field for the common parameter set, $a_0=0.25 $, $b_0=0.5$, $\omega=3.0$,  $\frac{k_BT}{m}=0.025$ and $\tau=0.1$. (a) $\frac{q\sqrt{D}}{m}=0.14$ (b) $\gamma_0=0.1$ (c) $\frac{q\sqrt{D}}{m}=0.274$ (d) $\gamma_0=0.85$. (Units are arbitrary.)}
\label{fig3}
\end{figure}

Understanding of the complex non-monotonic behavior of the rate constant may be helpful through the consideration of signature of the strengths of the damping and the fluctuating magnetic field on the barrier crossing dynamics. Fig.3 has been demonstrated in this context. Fig.3(a) shows that at the energy diffusion regime\cite{kram}, the first two turnovers merge  with the enhancement of the strength of the thermal noise due to increase in the damping strength.
Thus at relatively high damping strength the induced electric field may be effective in the dynamics even at low strength of the time independent magnetic field and as a result of that the tri-turnover becomes mono-turnover one. A similar situation may appear as demonstrated in Fig.3(b) for increase in the strength of the fluctuating magnetic field. But if the damping strength is large enough then all the turnover may disappear as shown in Fig.3(c). At high damping strength both the components of motion would lose  oscillating behavior and therefore the resonance phenomenon may not be observed. The motion along the $y$-direction also would be localized around the origin at this circumstance. Then the induced electric field may not be significant in the barrier crossing dynamics. This issue has been discussed in detail in Ref.\cite{physa}. Thus the resonance effect and the fluctuating magnetic field have important roles for the appearance of the tri-turnover phenomenon. 

Fig.3(d) shows that, on further increase in the strength of the fluctuating magnetic field at high damping strength, both the second and the third turnovers reappear. It clearly suggests that the first turnover is an interplay between the resonance effect (due to magnetic field induced coupling) and the decrease in the frequency factor of the barrier crossing rate with the increase in the strength of the time independent magnetic field. It also suggests that
the second turnover appears due to the leading role of the induced electric field. Finally, Fig.3(c) along with Fig.3(d) imply that the third turnover appears when the frequency factor becomes very low at high strength of the time independent magnetic field.

\begin{figure}[!htb]
\includegraphics[width=16cm,angle=0,clip]{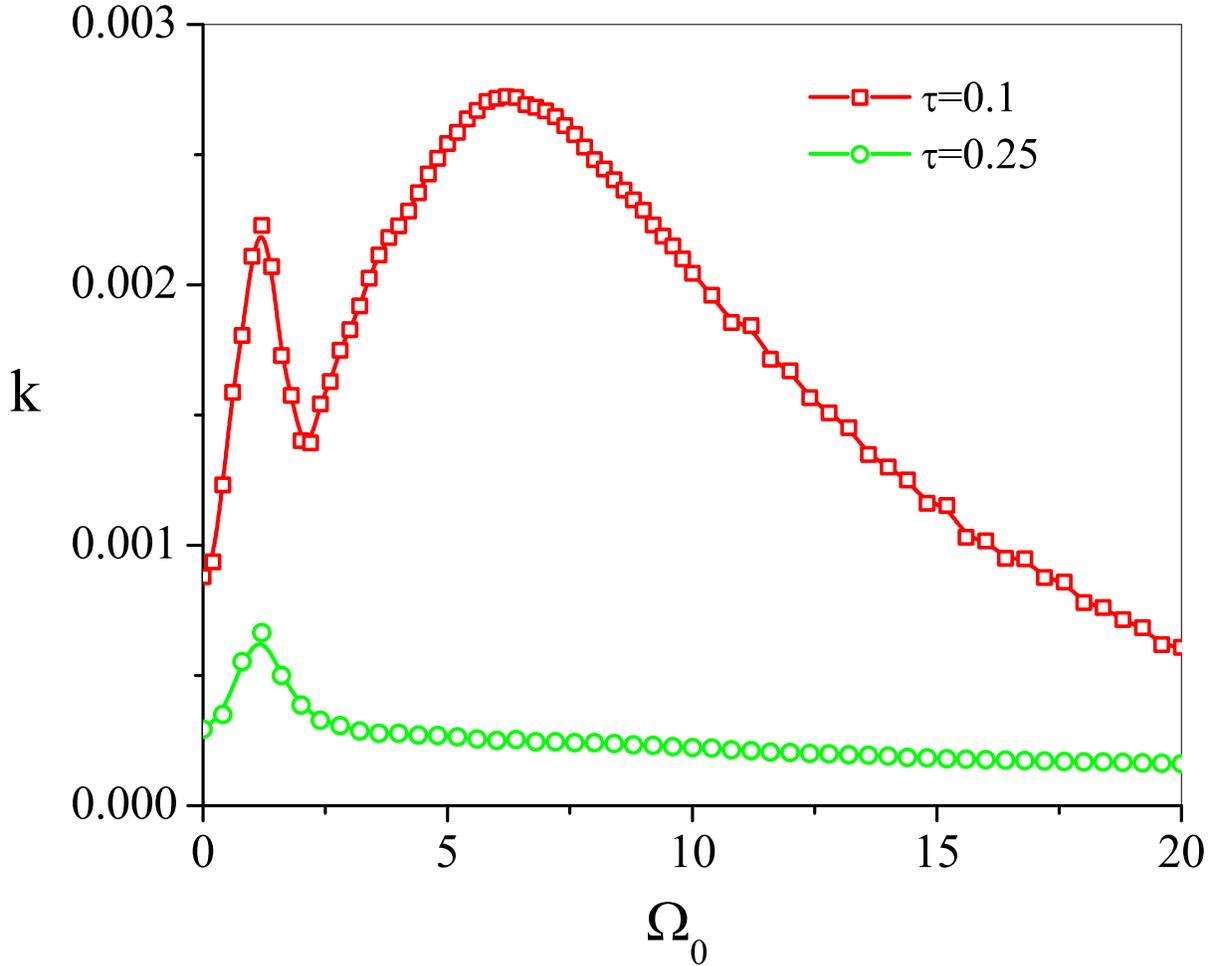}
\caption{(Color online) Conversion of tri turnover phenomenon into mono turnover one on changing correlation of the fluctuating magnetic field. 
Plot of the rate constant, $k$ {\it vs.} strength of the time independent magnetic field for the  parameter set, $a_0=0.25$, $b_0=0.5$, $\omega=3.0$, $\gamma_0=0.1$, $\frac{k_BT}{m}=0.025$ and  $\frac{q\sqrt{D}}{m}=0.14$. (Units are arbitrary.)}
\label{fig4}
\end{figure}

We now explore the effect of the correlation time of the fluctuating magnetic field on the non-monotonic behavior of the barrier crossing rate.
In this context 
we have presented the variation of the rate constant as a function of the strength of time independent magnetic field in Fig.4. It shows that at relatively large correlation time of the fluctuating magnetic field the tri-turnover phenomenon becomes a mono-turnover one with the disappearance of both the second and the third turnovers. 
This conversion can be understood easily based on the earlier discussion. The strength of the fluctuating magnetic field decreases with increase in the correlation time of the field as suggested by the two time correlation function of the fluctuating magnetic field
. Thus at relatively large noise correlation time, the induced electric field may be insignificant in the barrier crossing dynamics. Then the mono-turnover as shown in Fig.4 is a result of the interplay between 
the resonance effect and the decrease of the frequency factor of the rate constant with increase in the strength of the TIMF. At the same time this figure clearly implies that the second turnover appears when the induced electric field takes the leading role and the third turnover occurs if the frequency factor becomes very low at very large strength of the time independent magnetic field.

\begin{figure}[!htb]
\includegraphics[width=16cm,angle=0,clip]{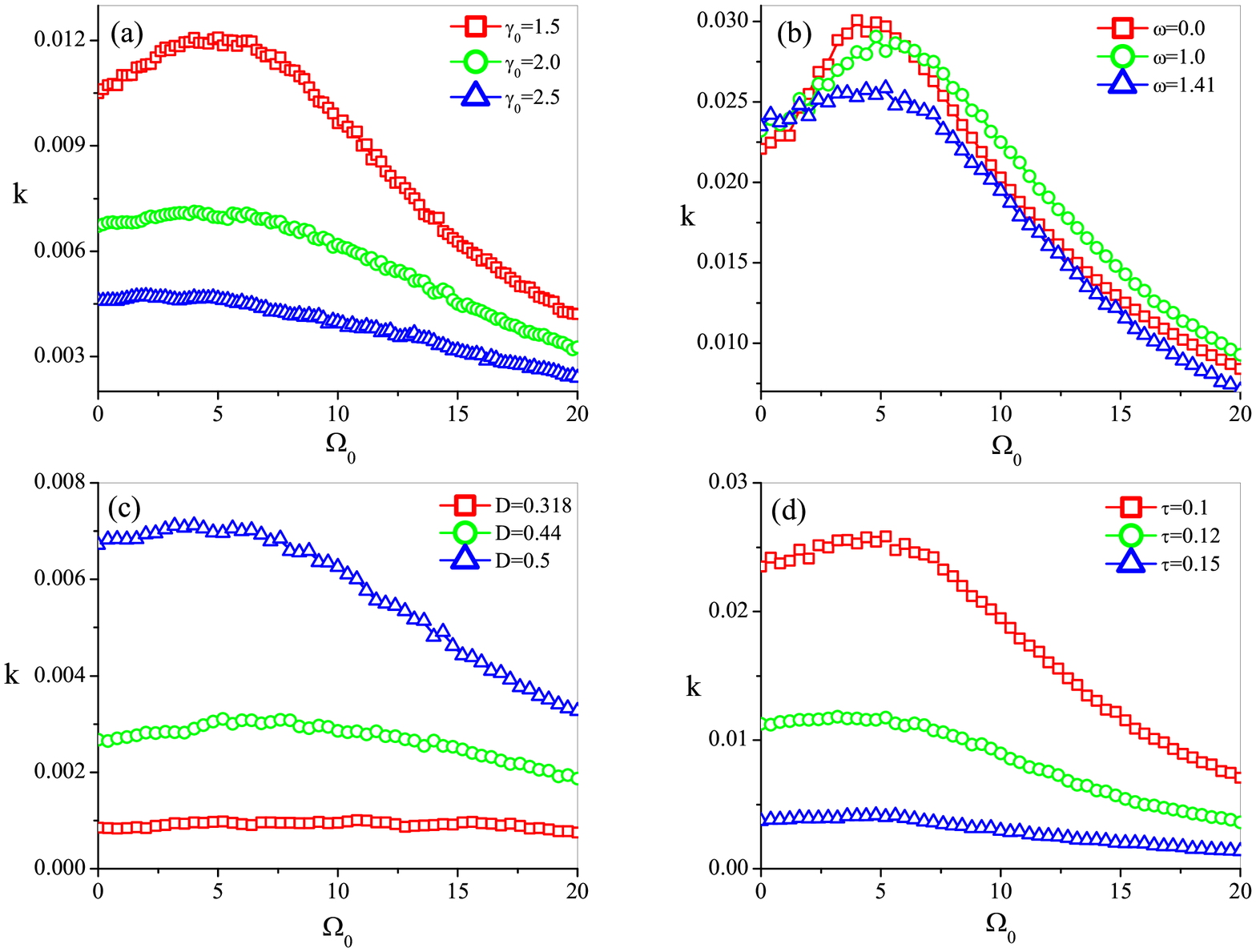}
\caption{(Color online) Fate of the tri turnover phenomenon on changing strengths of the damping. 
Plot of the rate constant, $k$ {\it vs.} strength of the time independent magnetic field for the common parameter set, $a_0=0.25 $, $b_0=0.5$,   $\frac{k_BT}{m}=0.025$  (a) $\omega=1.41$, $\frac{q\sqrt{D}}{m}=0.5$, $\tau=0.1$.(b) $\gamma_0=2.0$, $\frac{q\sqrt{D}}{m}=0.5$, $\tau=0.1$. (c) $\omega=1.41$, $\gamma_0=2.0$,  $\tau=0.1$. (d) $\omega=1.41$, $\gamma_0=2.0$,$\frac{q\sqrt{D}}{m}=0.5$. (Units are arbitrary.)}
\label{fig5}
\end{figure}

In the above discussion we have presented the mechanism of barrier crossing dynamics of a charged particle including the effect of the inertia.
But if the Brownian particle experiences a very high viscus force with large $\gamma_0$ then $\dot{u}_x$ may be negligible compared to $\gamma_0{u}_x$ in Eq. (\ref{eq4}). This is known as the over-damped approximation where $\gamma_0$ is higher compared to the frequency of the dynamical system. Under this circumstance  Eq. (\ref{eq4}) becomes\cite{aquino}

\begin{equation}
\gamma_0\dot{x} =-4 a_0 x^3+2 b_0 x +(\Omega_0+\Omega_f(t))\dot{y} +\frac{\dot{\Omega}_f(t) y}{2} + f_x(t)/m   \; \; \;. \label{eq4a}
\end{equation}

\noindent
Similarly, Eq. (\ref{eq5}) can be read as

\begin{equation}
\gamma_0\dot{y} = -\omega^2  y-(\Omega_0+\Omega_f(t))\dot{x} -\frac{\dot{\Omega}_f(t) x}{2} +f_y(t)/m \; \; \;.    \label{eq5a}
\end{equation}
\noindent

Solving the coupled equations (\ref{eq12f}, \ref{eq4a}-\ref{eq5a}) we have calculated the barrier crossing rate constant as a function of the strength of the time independent magnetic field and plotted in Fig.5. Here we have chosen that $\gamma_0$ is higher compared to $\omega_a$, $\omega_b$ and $\omega$, respectively. Then frequency of the dynamical system is certainly lower than $\gamma_0$ for entire range of $\Omega_0$
as suggested by the above analysis. However, 
it is clear in in Fig.5 that the first turnover does not appear if the effect of inertia is not important in the dynamics. This happens because at this situation both the components of motion may lose the oscillating behavior which signifies the absence of the resonance effect as well the first turnover. It is also clear in Fig.5 that at relatively high damping strength, even the second turnover may disappear as shown in panel (a) of Fig.5. Thus induced electric field may lose its effectiveness in the dynamics due to the localization of the motion around the origin along $y$-direction at this condition. The localization of the motion may also occur at high frequency and then the rate constant may decrease monotonically as shown in panel (b) of the same figure. Similarly the rate constant may decrease regularly (as implied in panels (c) and (d) of Fig.5) either at low strength of the fluctuation magnetic field or high correlation time of the filed since the variance of the random field is small in both cases. Thus the mechanism of the barrier crossing dynamics in the presence of a magnetic field is relatively simple if the effect of the inertia is not important in the dynamics.

\begin{figure}[!htb]
\includegraphics[width=16cm,angle=0,clip]{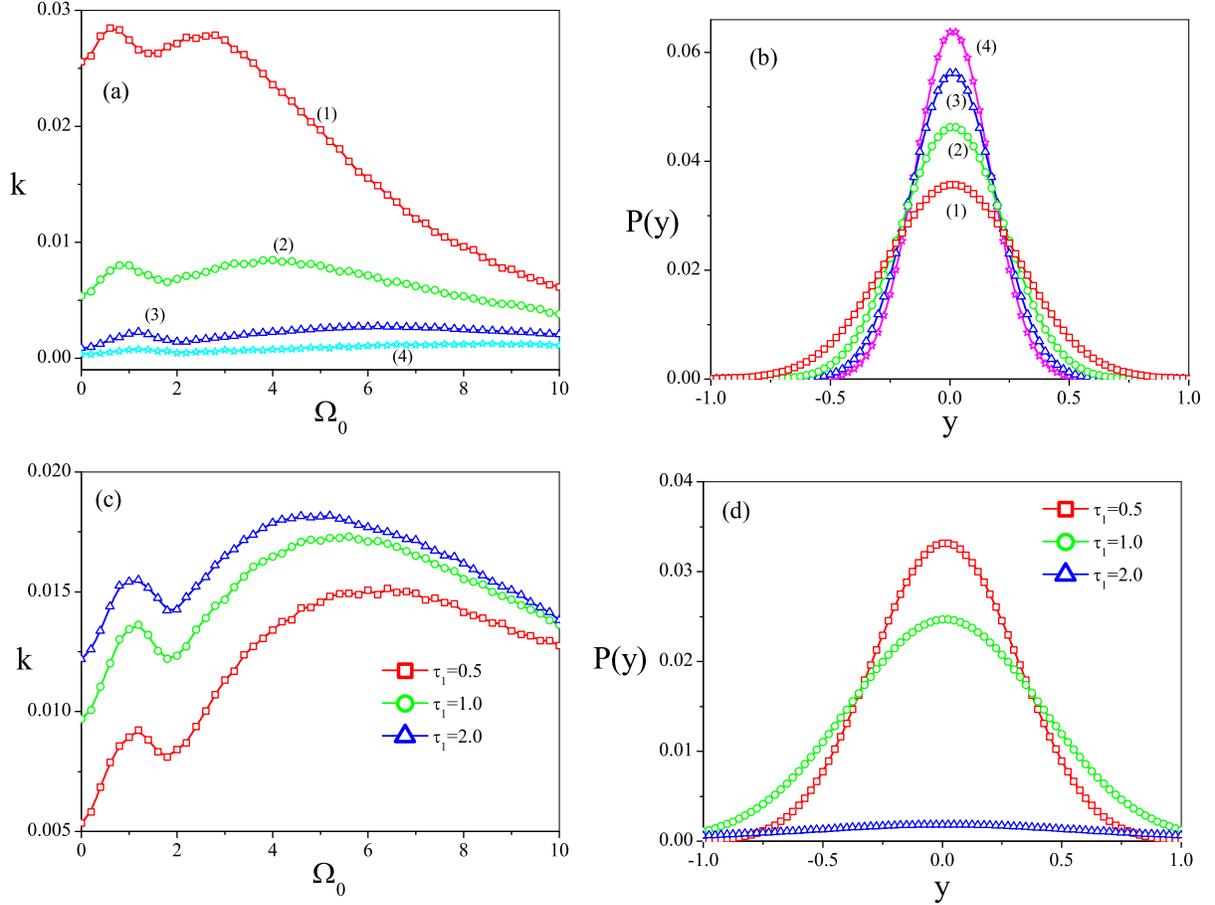}
\caption{(Color online) The effect of the mass and the correlation time of the thermal noise on the fluctuating magnetic field induced tri-turnover phenomenon.  The common parameter set, $\omega=3.0$,  , $\gamma_0=0.1$, and $\tau=0.1$.
(a) Plot of the rate constant, $k$ {\it vs.} strength of the time independent magnetic field for different mass of the Brownian particle and parameter set(1) $a_0=0.5$, $b_0=1.0$, $\frac{k_BT}{m}=0.05$ and $\frac{q\sqrt{D}}{m}=0.28$  (2) $a_0=0.33$, $b_0=0.67$, $\frac{k_BT}{m}=0.033 $ and $\frac{q\sqrt{D}}{m}=0.188$,(3) $a_0=0.25$, $b_0=0.5$, $\frac{k_BT}{m}=0.025$ and $\frac{q\sqrt{D}}{m}=0.14$ and (4)$a_0=0.20$, $b_0=0.41 $, $\frac{k_BT}{m}=0.020$ and $\frac{q\sqrt{D}}{m}=0.118$  (b) Plot of reduced distribution function, $P(y)$ {\it vs.} $y$ for the same parameter set as in (a).
(c) Plot of the rate constant, $k$ {\it vs.} strength of the TIMF for different correlation time of the thermal noise for the parameter set, $a_0=0.25$ and $b_0=0.25$, $\omega=3.0$, $\frac{k_BT}{m}=0.025$ and $\frac{q\sqrt{D}}{m}=0.14$ $\gamma_0=0.1$, and $\tau=0.1$. (d) Plot of reduced distribution function, $P(y)$ {\it vs.} $y$ for the same parameter set as in (c). (Units are arbitrary.)}
\label{fig6}
\end{figure}

We now examine the influence of the mass of the Brownian particle on the non-monotonic behavior of the rate constant.
The terms with the magnetic force and the induced electric field in the Langevin Eqs.(\ref{eq14}-\ref{eq15}) are only mass independent. Effectiveness of these terms  may be influenced by the mass of the Brownian particle in the barrier crossing dynamics. To explore this aspect we have calculated the rate constant for different values of $m$ and plotted these in Fig.\ref{fig6}(a). For this calculation we have kept fixed the curvature of the double well potential as well as the activation energy to identify the fluctuating electric field induced activation of the Brownian particle. Fig.\ref{fig6}(a) shows that the tri-turnover may become a mono-turnover one at relatively low mass. The disappearance of the second turnover means that at this limit the induced electric field may take a leading role for the activation of the Brownian particle  even at low strength of the time independent magnetic field. For, the low mass enhancement of the acceleration as well as width of distribution function in terms of coordinate of $y$-direction (as shown in Fig.6(b)) facilitates the induced electric field to take the leading role. Thus the fluctuating magnetic field has an important role for the appearance of the tri-turnover phenomenon.

Finally, we examine the effect of the non-Markovian thermal noise on the non-monotonic behavior of the rate constant.
For the non-Markovian thermal bath the equations of motion (\ref{eq14}-\ref{eq15}) become\cite{physa}

\begin{equation}
\dot{u}_x =-4 a_0 x^3+2 b_0 x -\int_0^tdt'\gamma(t-t')\dot{x}(t')+(\Omega_0+\Omega_f(t)) u_y-\frac{\Omega_f(t)y}{2\tau}+\frac{q\sqrt{D}y}{2q\tau} \zeta(t) + f_x(t)/m   \; \; \; \label{eq35a}
\end{equation}

\noindent
and

\begin{equation}
\dot{u}_y = -\omega^2  y-\int_0^tdt'\gamma(t-t')\dot{x}(t')-(\Omega_0+\Omega_f(t)) u_x +\frac{\Omega_f(t)}{2\tau}-\frac{q\sqrt{D}x}{2m\tau} \zeta(t) +f_y(t)/m   \; \; \;, \label{eq36}
\end{equation}

\noindent
where the time dependent damping strength is related to the thermal noise by the relation,
$\langle f_x(t) f_x(t')\rangle=  \langle f_y(t)f_y(t')\rangle=\langle f_z(t)f_z(t')\rangle=k_BTm\gamma(t-t')$. 
To capture the essential
features of the non-Markovian dynamics, we consider Drude-Ullersma model \cite{sraypre}
for the frequency dependence of the coupling coefficients which gives the exponentially decaying frictional
memory kernel \cite{alendu_jcp,marjcp,Okuyama_JCP86,sraypre,sraypre1}.  
Then $\gamma(t-t')$ in Eqs.(\ref{eq35a},\ref{eq36}) can be read as,
$\gamma(t-t^{\prime}) = \frac{\gamma_0}{\tau_1}\exp\left(\frac{-|t-t^{\prime}|}{\tau_1}\right)$. Thus $\tau_1$ is correlation time of the thermal noise. With this form of the time dependent damping strength we have solved
Eqs.(\ref{eq1}-\ref{eq2},\ref{eq12},\ref{eq35a}) and (\ref{eq36}) and  plotted the rate constant {\it vs.} $\Omega_0$ in Fig.\ref{fig6}(c) for different $\tau_1$. 
It is apparent in this figure that the induced electric field becomes more significant in the dynamics with increase in noise correlation time of the thermal noise. As the thermal bath deviates more from the Markovian character the bath modes of high frequencies are ignored\cite{sraypre1}. 
Then the system experiences a less damping strength and as a result of that the width of $P(y)$ is enhanced (as shown in Fig.6(d)) with increase in     
correlation time of the thermal noise. Thus the induced electric field becomes more effective in the activated barrier crossing process with increase in the correlation time of the thermal noise.

\section{conclusion}
We have studied the barrier crossing dynamics of a Brownian particle in the presence of both the time independent and the fluctuating magnetic fields. Our investigation clearly explores the mechanism of the activated barrier crossing process. It includes the following major points:

(i) We have explored the origin of the appearance of a turnover in the variation of the rate constant as a function of strength of the time independent magnetic field. It may disappear for the two cases. (a) Frequency of the harmonics oscillator along the $y$ direction is far different from the frequency at the bottom of the double well oscillator. (b) Strength of the damping is relatively high. These observations imply that a resonance effect may appear due to the magnetic field induced coupling. We have solved the equation of motion in the absence of the fluctuating magnetic field and the thermal bath. The solution describes a periodic motion with the same frequency for both the directions when the frequency of the harmonics oscillator along the $y$ direction matches with the frequency at the bottom of the double well oscillator. But the frequency of the periodic motion decreases with increase in the strength of the time independent magnetic field. Thus the turnover appears as a result of the interplay between the resonance effect and the decrease of the frequency factor of the rate constant. 
It is corroborated by the fact that the turnover disappears in the absence of the harmonic force along the $y$ direction.

(ii) In the presence of fluctuating magnetic field two additional turnovers may appear at relatively large strength of the time independent magnetic field. Another interesting observation is that the first turnover disappears in the absence of the harmonic force along the $y$ direction. Thus the first turnover is due to the interplay as mentioned above. Another significance of the observation in the absence of the harmonic force is that the motion may be bounded in space by virtue of the fluctuating induced electric field.

(iii) The energy dissipation may become weak with increase in the strength of the TIMF since  the effective frequency of the motion decreases as the field intensity grows. Thus the effectiveness of the coordinate dependent induced electric field (which activates the Brownian particle) becomes more significant as the strength of the time independent magnetic field is enhanced. Hence at relatively high strength of the TIMF, the induced electric field may take a leading role in the barrier crossing dynamics and then the second turnover may appear. Finally, very low frequency factor at very high strength of the time independent magnetic field may introduce the third turnover. Thus the second and the third turnovers are the result of the interplay between the fluctuating electric field induced activation and the decrease in the frequency factor with increase in the strength of the time independent magnetic field. This has been corroborated by the following facts. At relatively large strength of the fluctuating magnetic field only single turnover appears as a result of merging of the first and the second turnovers. But the single turnover disappears at over damped limit. Under this circumstance, if we increase the strength of the FMF then the second and the third turnovers appear. It certainly implies that these turnovers are due to the interplay between the fluctuating electric field induced activation and the decrease of the frequency factor with increase in the strength of the time independent magnetic field.  

Another important point is to be noted here. It is apparent in the above discussion that the second and the third turnovers are due to the fluctuating magnetic field. But in the presence of the fluctuating magnetic field the resonance effect may appear even for the case where the frequency of the harmonic potential along $y$-direction is far different from the frequency at the bottom of the double well potential. In that case the tri-turnover phenomenon is a purely fluctuating magnetic field induced one. Thus the coordinate dependent
fluctuating induced electric field may offer the effective required frequencies to have the resonance effect.

(iv) The tri-turnover phenomenon may be transformed into a mono-turnover one at relatively large correlation time of the fluctuating magnetic field. At this limit the fluctuating magnetic field may be insignificant in the activated barrier crossing process since its variance becomes very small. Thus the second and the third turnovers are clearly due to the interplay between the fluctuating electric field induced activation and  the decrease of the frequency factor of the rate constant with increase in the strength of time independent magnetic field.

(v) If the effect of the inertia is not important in the dynamics then the mechanism of the barrier crossing dynamics in the presence of a magnetic field is relatively simple.

In addition to the tuning of the conductivity of the solid electrolyte as mentioned in the introduction section,
the present study may be relevant in the following situations. In the several recent experiments\cite{hara,sun} the influence of the magnetic field on the transport of ion has been investigated. To study the ion transport of cultured cells both time dependent and independent magnetic fields have been considered in Ref.\cite{hara}. Another very recent study\cite{sun} where the ion migration process is modulated with the magnetic field.  
In this context one may also consider the time dependent magnetic field. Then we would mention the following point. 
It has been reported that ion moves in the condensed phase through hopping mechanism \cite{rob,con} i.e., activated barrier crossing process. Thus the present study may be helpful to account for experimental observations on the ion transport in the presence of a time dependent magnetic field. Finally, the calculation of the frequency of the deterministic dynamics in the presence of the time independent magnetic field
implies that one may tune the barrier crossing rate by the proper choice of the electric field which is periodic in time. 
Studies on this aspect and the effect of magnetic field on autonomous and non autonomous stochastic resonances may appear shortly.

\vspace{0.5cm}

\noindent
{\bf Acknowledgment}

\noindent
Thanks are due to 
We are thankful to Prof. Abhijit Sen for critically reading the manuscript. S. M. is thankful to Council of Scientific and Industrial Research for partial financial support.

\appendix
\section{•}

In the presence of fluctuating magnetic field ($\eta(t)$) Eq.(\ref{eqzeta}) becomes

\begin{equation}\label{A1}
\ddot{\xi}=-\alpha \xi-\beta \dot{\xi} \; \;,
\end{equation}
\noindent
where $\alpha=(\omega^2+i\frac{\dot{\Omega}_f(t)}{2})$ and $\beta= \gamma_0+i(\Omega_0+\Omega_f)$. At over damped limit it reduces to

\begin{equation}
\dot{\xi}=A(t)\xi \; \;,
\end{equation}
\noindent
where $A(t)=\frac{\alpha}{\beta}$. Its solution can read as

\begin{equation}
\xi=\xi_0 e^{-\int_0^tA(t')dt'}
\end{equation}

Now identifying real and imaginary parts of $A(t)$ with

\begin{equation}
A_R=\frac{\omega^2 \gamma_0 + (\Omega_0+\Omega_f(t)) \dot{\Omega}_f(t)/2}{\gamma_0^2+(\Omega_0+\Omega_f(t))^2}
\end{equation}

\noindent
and

\begin{equation}
A_I=\frac{\gamma_0 \dot{\Omega}_f(t)/2-\omega^2(\Omega_0+\Omega_f(t))}{\gamma_0^2+(\Omega_0+\Omega_f(t))^2} \; \;,
\end{equation}

\noindent
one can read real and imaginary parts of $\xi$ as

\begin{equation}
x(t)= x_0 \cos (\theta) e^{-\int_0^t A_R(t')dt'}+y_0 \sin(\theta) e^{-\int_0^t A_R(t')dt'}
\end{equation}

\noindent
and

\begin{equation}
y(t)= y_0 \cos (\theta) e^{-\int_0^t A_R(t')dt'}-x_0 \sin(\theta) e^{-\int_0^t A_R(t')dt'} \; \;.
\end{equation}
\noindent
Here $\theta =\int_0^t A_I(t')dt'$.


\begin{thebibliography}{99}
\bibitem{scros} B. Scrosati (ed.), {\it Applications of Electroactive Polymers} (Chapman \& Hall, London,1993);
P. G. Bruce, {\it Solid State Electrochemistry} (Cambridge Univ. Press, Cambridge,1995);
F. M. Gray, {\it Polymer Electrolytes} (RSC Materials Monographs, The Royal Society of Chemistry, Cambridge, 1997);
J. M. Tarascon  and  M. Armand, Nature (London) {\bf 414},359 (2001);                                                
G.S.MacGlashan, Y. G. Andreev  and P. G. Bruce, Nature (London) {\bf 398}, 792 (1999);
A.Barnes, A. Despotakis, T. C. P. Wong, A. P. Anderson, B. Chambers, and P. V. Wright,  Smart Mater. Struct. {\bf 7}, 752 (1998). 

\bibitem{angel} C. A.  Angell, C. Liu, and E. Sanchez,  Nature (London) {\bf 137}, 362( 1993);
A. M. Christie, S. J. Lilley, E. Staunton, Y. G. Andreev and P. G. Bruce, Nature (London)  {\bf 433} 50(2005). 


\bibitem{katsuki} K. Amemiya, J. Phys. Soc. Jpn. {\bf 72}, 135 (2003).

%Amemiya K 2003 {\it J. Phys. Soc. Jpn.} {\bf 72} 135


\bibitem{pere} A. S. Moskalenkoa, S. D. Ganicheva, V. I. Pere\'{l}  and I.N. Yassievicha,  Physica B {\bf 273-274}, 1007 (1999);
V. I. Pere\'{l}  and I.N. Yassievich, JETP Letts. {\bf 68}, 804 (1998);
A. S. Moskalenko, V. I. Pere\'{l}  and I. N. Yassievich, JETP {\bf 90}, 217 (2000). 

\bibitem{telang} N. Telang  and S. Bandyopadhyay, Appl. Phys. Letts. {\bf 66}, 1623 (1995). 

\bibitem{vdo} E. E. Vdovin, A. Levin, A. Patan\`{e}, L. Eaves, P. C. Main, Y.N. Khanin, Y.V.Dubrovskii, M. Henini and G. Hill, Science {\bf 290}, 122 (2000). 


\bibitem{bag3} A. Baura, M. K. Sen and B. C. Bag,  Phys. Chem. Chem. Phys. {\bf 13}, 9445 (2011). 


\bibitem{bag4} A. Baura, M. K. Sen and B. C. Bag, Chem.  Phys. {\bf 417}, 30 (2013). 


\bibitem{aquino} J. I. Jim\'enez-Aquino and M. Romero-Bastida,  Phys. Rev. E {\bf 86}, 031110 (2012).


\bibitem{bag5}A. Baura, S. Ray and B. C. Bag, J. Chem. Phys. \textbf{138}, 244110 (2013).

\bibitem{bag6} S. Mondal, S. Das, A. Baura and B. C.Bag, J. Chem. Phys. \textbf{141}, 224101 (2014). 


\bibitem{physa} S. Mondal, A. Baura, S. Das and B. C. Bag , Physica A {\bf 502}, 58 (2018). 

\bibitem{spre} S. Mondal and B. C. Bag, Phys. Rev. E {\bf 91}, 042145 (2015).

\bibitem{indmag}M. L. Mittal, Y. S. Prahalad   and D. G. Thirtha,  J. Phys. A: Math. Gen. {\bf 13}, 1095 (1980).
A.Saha   and  A. M. Jayannavar, Phys. Rev. E {\bf 77}, 022105(2008 ).


\bibitem{bagmal} Th. Leiber, F. Marchesoni and H. Risken,  Phys. Rev. Lett. {\bf 59}, 1381(1987);
P.Jung, P.  H\"{a}nggi, F. Marchesoni,  Phys. Rev. A {\bf 40}, 5447 (1989);
S. K.Banik, J. R. Chaudhuri and D. S. Ray, J. Chem. Phys. {\bf 112}, 8330 (2000);
B. C. Bag, Phys. Rev. E {\bf 66}, 026122 (2002).

\bibitem{hang1} P. H\"{a}nggi  and P. Jung, Adv. Chem. Phys. {\bf 89}, 239 (1995).


\bibitem{rob}R. A. Robinson  and R. H.Stokes,  {\it Electrolytic Solutions} (Butterworth, London, 1955);
I. Rubinstein, M. Bixon  and E. Gileadi, J. Phys. Chem. {\bf 84}, 715 (1980). 

\bibitem{con}B. E. Conway, J. O'M. Bockris and H. Linton, J. Chem. Phys. {\bf 24}, 834( 1956).

\bibitem{christie}A. M.  Christie, S. J. Lilley, E. Staunton, Y. G. Andreev and P. G. Bruce, Nature (London) {\bf 433}, 50 (2005).

\bibitem{muru}S. Murugavel, C. Vaid, V. S. Bhadram and C. Narayana, J. Phys. Chem. B {\bf 114}, 13381(2010). 

\bibitem{strom} U. Strom, M. Von Schickfuss  and S. Hunklinger, Phys. Rev. B {\bf 25}, 2405 (1982); 
U. Strom, Solid State Ionics {\bf 8}, 255 (1983);
K. Funke, Prog. Solid St. Chem. {\bf 22}, 111 (1993);
M. W. Breiter, W. J. Lorenz  and  G. Staikov, Solid State Phenomena {\bf 39-40}, 3 (1994);
 S.Murugavel, C.  Vaid, V. S. Bhadram  and  C. Narayana, J. Phys. Chem. B {\bf 114}, 13381 (2010). 

\bibitem{kram}H. A.  Kramers, Physica (Utrecht)  {\bf 7}, 284 (1940).

\bibitem{hangimoj}P. H\"{a}nggi  and F. Mojtabai, Phys. Rev. A {\bf 26}, 1168 (1982).

\bibitem{marche} F. Marchesoni, Adv. Chem. Phys. {\bf 63}, 603 (1985).

\bibitem{revm} P. H\"{a}nggi, P. Talkner and M. Borkovec , Rev. Mod. Phys. {\bf 62}, 251 (1990). 

\bibitem{melni} V. I. Mel'nikov, Phys. Rep. {\bf 209}, 1 (1991).

\bibitem{revm1}L. Gammaitoni, P. H\"anggi, P. Jung, and F. Marchesoni, Rev. Mod. Phys. {\bf 70} ,223 (1998). 

\bibitem{jsm}M. K. Sen, A.Baura and B. C.Bag, J. Stat. Mech.,  P11004 (2009).

%\bibitem{garcia1}
%P. L. García-Müller, R. Hernandez, R. M. Benito, and F. Borondo, J. Chem.
% Phys. {\bf 137}, 204301 (2012);  J. J. Mazo, O. Y. Fajardo, and D Zueco,  J. Chem. Phys.  {\bf 138}, 104105 (2013).
%P. Tiwary and M. Parrinello, Phys. Rev. Lett. {\bf 111}, 230602 (2013)

\bibitem{alendu_jcp} A. Baura, M. K. Sen, G. Goswami and B. C. Bag, J. Chem.  Phys. {\bf 134}, 044126( 2011); S.Ray, D. Mondal and  B. C. Bag , J. Chem. Phys. {\bf 140}, 204105 (2014). 

%\bibitem{guan}
%R. Guantes, J. L. Vega, S. Miret-Artes and E.  Pollak,  J. Chem. Phys. {\bf 119}, 2780 (2003); 
%R. B. Best and G. Hummer, Phys. Rev. Lett. 96, 228104 (2006).                                           ́

%\bibitem{poll2}
%E.  Pollak and J. Ankerhold, J. Chem. Phys. {\bf 138}, 164116 (2013).

\bibitem{poll3} E.  Pollak and R. Ianconescu, J. Chem. Phys. {\bf 140}, 154108 (2014). 

%\bibitem{garcia}
%P. L. Garcia-Muller, R. Hernandez, R. M. Benito and F. Borondo,  Phys. Rev. Lett. 
%{\bf 137}, 204301 (2012);
%J. Wu, R. J. Silbey, and J. Cao, Phys. Rev. Lett. 110, 200402 (2013).

\bibitem{poll5} R. Ianconescu  and E. Pollak, J. Chem. Phys. {\bf 143}, 104104 (2015). 

\bibitem{tiwar}
P.Tiwary  and B. J. Berne, J. Chem. Phys. {\bf 144}, 134103 (2016). 

\bibitem{jsm1} S. Mondal, B. C. Gupta  and B. C. Bag, J. Stat. Mech., 013204 (2016).

\bibitem{chaos} Y. Li and X. Liu, Chaos {\bf 29}, 023137 (2019)


\bibitem{chaos1} M. A. H\"Ogele and I. Pavlyukevich, Chaos {\bf 29}, 063104 (2019).

%\bibitem{poll6}
%E. Pollak and R. Ianconescu, J. Phys.  Chem. A {\bf 120}, 3155 (2016).


%\bibitem{khoo}K. L. Khoo, L. A. Dissado, J. Fothergill, and  I. J. Youngs,
%{\it Proceedings of the 2004 IEEE International Conference on Solid Dielectrics},
%{\bf 2}, 550 (2004).

%\bibitem{mason} D. R. Mason, J. le. Page, C. P. Race, W. M. C.  Foulkes, M. W. Finnis and A. P.
%Sutton, J. Phys.: Condens. Matter {\bf 19}, 436209 (2007). 


%\bibitem{khoo1}K. L. Khoo  and  L. A.Dissado, {\it Molecular 
%Dynamics – Theoretical Developments and Applications in Nanotechnology and Energy}, InTech,  371 (2012) 


\bibitem{aquino1}J. I. Jim\'enez-Aquino, R. M. Velasco  and  F. J. Uribe, Phys. Rev.E {\bf 77}, 051105 (2008).  

\bibitem{hang}P. H\"anggi, F. Marchesoni and  P.Grigolini, Z. Phys. B {\bf 56}, 333 (1984).

\bibitem{jia1}Y. Jia ,S. Yu , and J. Li ,  J, Phys. Rev. E {\bf 62}, 1869 (2000).  

\bibitem{cao}
C. Li, W. Da-jin and K. Sheng-zhi, Phys. Rev. E  {\bf 52}, 3228 (1995).  

\bibitem{rog} R. S. M. Rikken, R. J. M. Nolte, J. C. Maan, J. C. M. van Hest, D. A. Wilson and  P. C. M. Christianen, Soft Matter {\bf 10}, 1295 (2014).

\bibitem{levine} I. N. Levine   {\it Physical Chemistry} (McGraw-Hill Higher Education; 1st September, 2001).


\bibitem{vice}J. de Vicente, J. P. Segovia-Gutierrez, E. Andablo-Reyes, F.  Vereda and R. Hidalgo-Alvarez, J. Chem. Phys. {\bf 131}, 194902 (2009). 

\bibitem{hsieh} T. H. Hsieh  and H. J. Keh, J. Chem. Phys. {\bf 138}, 0741051 (2013). 

\bibitem{pnik} R. Czopnik and P.Garbaczewski, Phys. Rev. E {\bf 63}, 021105 (2001). 

\bibitem{mar} J. I. Jim\'{e}nez-Aquino and M. Romero-Batida, Phys. Rev. E {\bf 76}, 021106 (2007);
A. M.Jayannavar  and  M. Sahoo, Phys. Rev. E {\bf 75}, 032102(2007);
A. Saha and A. M. Jayannavar, Phys. Rev. E {\bf 77}, 022105 (2008);  
D. Roy  and N. Kumar, Phys. Rev.E {\bf 78}, 052102 (2008);
A. Baura, M. K. Sen and B. C. Bag, Phys. Rev. E {\bf 82}, 041102 (2010).

\bibitem{fabio} S. E.  Savel\'{e}v and F. Marchesoni, Phys. Rev. E {\bf 90}, 062117 (2015). 

\bibitem{jayn} P. S. Pal, S. Rana, A. Saha, and A. M. Jayannavar, Phys. Rev. E {\bf 90}, 022143 (2014);
T. Chen, X. B. Wang  and T. Yu, Phys. Rev. E {\bf 90}, 022101 (2014);
S. Ray, M. Rano and B C Bag, J. Chem. Phys. {\bf 142}, 154122 (2015). 

\bibitem{gelf}A. K. Ram and B. Dasgupta, Physics of Plasmas {\bf 17}, 122104 (2010 );
V. Gelfreich, V. Rom-Kedar, K. Shah and D. Turaev, Phys. Rev. Letts. {\bf 106}, 074101 (2011). 

\bibitem{nmar}F. N. C.Paraan, M. P. Solon, and J. P. Esguerra, Phys. Rev. E {\bf 77}, 022101 (2008). 

\bibitem{nmar1}A. Baura, M. K. Sen and B. C. Bag,  Eur. Phys. J. B {\bf 75}, 267 (2010);
A. Baura, S. Ray, M. K.Sen and B. C. Bag, J. Appl. Phys. {\bf 113}, 124905 (2013).  

\bibitem{nmar3}J. I.Jim\'{e}nez-Aquino, Phys. Rev. E {\bf 92}, 022149 (2015). 

\bibitem{nmar4}J. C. Hidalgo-Gonzalez, J. I. Jim\'{e}nez-Aquino  and M. Romero-Bastida, Physica A {\bf 462}, 1128 (2016). 

\bibitem{jcp} J. Das, S. Mondal and B. C. Bag, J. Chem. Phys. {\bf 147}, 164102 (2017).

\bibitem{rt} R. Toral, {\it In Computational Physics, Lecture Notes in Physics, vol. 448} (Springer-Verlag, Berlin, 1995).

\bibitem{landau} L. D. Landau  and E. M. Lifshitz ,{\it The Classical Theory of Fields} (Third Revised English Edition, Pergamon Press Ltd.,  Oxford, 1971)

\bibitem{sraypre} S. Ray  and B. C. Bag, Phys. Rev. E {\bf 92}, 052121 (2015).   

\bibitem{marjcp} F. Marchesoni  and P.Grigolini, J. Chem. Phys. {\bf 78}, 6287 (1983). 

\bibitem{Okuyama_JCP86} S. Okuyama and D. W. Oxtoby, J. Chem. Phys. {\bf 84}, 5830 (1986). 

\bibitem{sraypre1} S. Ray and B. C.Bag, Phys. Rev. E {\bf 90}, 032103 (2014). 


%\bibitem{srayjcp} S. Ray, D. Mondal and B. C. Bag,  J. Chem. Phys. {\bf 140}, 204105 (2014).

\bibitem{hara}T. Ikehara ,H. Yamaguchi and H.Miyamoto, J. Med. Invest. {\bf 45}, 47 (1998).  

\bibitem{sun}P. Sun ,F. Zheng  ,K. Wang ,M. Zhong ,D. Wu and H. Zhu , Sci. Rep. {\bf 4},  6798( 2014).
\end{thebibliography}
\end{document}